\documentclass[aps,amsmath,prl,twocolumn,nofootinbib]{revtex4}

\usepackage{amsmath}
\usepackage{amssymb}
\usepackage{graphicx}
\usepackage{hyperref}
\usepackage{epstopdf}

\newcommand{\abs}[1]{\lvert{#1}\rvert}
\newcommand{\avg}[1]{\langle{#1}\rangle}
\newcommand{\diff}[2]{\frac{d{#1}}{d{#2}}}

\newcommand{\pdiff}[2]{\frac{\partial{#1}}{\partial{#2}}}
\newcommand{\inlinepdiff}[2]{\partial_{#2}{#1}}

\newcommand{\inlinepdiffn}[3]{\partial_{#2}^{#3}{#1}}
\newcommand{\mm}[0]{\mu{m}}

\begin{document}
\date{\today}

\title{Signalling noise enhances chemotactic drift of \emph{E.~coli}}
\author{Marlo Flores}
\affiliation{FOM Institute AMOLF, Science Park 104, 1098XE, Amsterdam, Netherlands}
\author{Thomas S. Shimizu}
\affiliation{FOM Institute AMOLF, Science Park 104, 1098XE, Amsterdam, Netherlands}
\author{Pieter Rein ten Wolde}
\affiliation{FOM Institute AMOLF, Science Park 104, 1098XE, Amsterdam, Netherlands}
\author{Filipe Tostevin}
\affiliation{FOM Institute AMOLF, Science Park 104, 1098XE, Amsterdam, Netherlands}

\begin{abstract}
Noise in transduction of chemotactic stimuli to the flagellar motor of 
{\em E.~coli} will affect the random run-and-tumble motion of the cell and the
ability to perform chemotaxis. Here we use numerical simulations to show that an
intermediate level of noise in the slow methylation dynamics enhances drift
while not compromising localisation near concentration peaks. A minimal model
shows how such an optimal noise level arises from the interplay of noise and the
dependence of the motor response on the network output. Our results suggest that
cells can exploit noise to improve chemotactic performance.
\end{abstract}

\maketitle

The motion of \emph{Escherichia~coli} consists of a series of ``runs'', where
the cell swims in a roughly constant direction, and ``tumbles'', during which
the cell randomly reorients \cite{Berg72}. In a spatially-varying environment
the bacterium performs chemotaxis by biasing this random motion in the direction
of favourable conditions. A well-studied signalling cascade \cite{Berg_book}
detects environmental ligand stimuli and regulates the stochastic switching of
the rotary flagellar motors between counter-clockwise (run) and clockwise
(tumble) rotation \cite{Berg72}. However, the biochemical reactions making up
this signalling pathway are inherently random events, and the noise introduced
to the signal in this way will affect the swimming behaviour and ability of a
cell to respond to gradient stimuli. Here we show how signalling noise can be
beneficial for chemotactic performance.

Chemoattractant stimuli are detected by the binding of ligands to membrane
receptor complexes, which suppresses the activity of the receptor-associated
kinase CheA. Consequently, the phosphorylation level of the response regulator
CheY, which in its phosphorylated form (CheYp) binds to the flagellar motor and
promotes tumbling, decreases leading to longer runs. Conversely, a reduction in
ligand binding leads to an increase in CheYp and shorter runs. Hence, the random
walk motion of the bacterium is biased in the direction of increasing
chemoattractant concentration. Crucially, the chemotaxis network also includes a
negative feedback from CheA activity to receptor methylation. This ensures
adaptation of the network response to constant stimuli, enabling the network to
detect temporal derivatives of the observed ligand signal \cite{Segall86} and
allowing sensitivity to a wide range of ligand concentrations \cite{Sourjik02}. 

Receptor methylation and demethylation reactions are a significant source of
noise in the signalling network \cite{Korobkova04}, because (i) the timescale of
methylation ($\sim10s$ \cite{Shimizu10}) is much longer than the other
timescales in the network (ligand-receptor binding and receptor activity
changes: $\sim1ms$; CheY phosphorylation and motor switching: $\lesssim1s$),
meaning that the downstream network cannot integrate out slow methylation
fluctuations; and (ii) methylation occurs at a small number of sites on each
receptor catalysed by a small number of the enzymes CheR and CheB, meaning that
small-number fluctuations in the overall methylation level can be significant.
Importantly, the output of the noisy signalling network also affects the ligand
signal experienced by the cell via modulation of the tumbling dynamics and hence
the swimming trajectory. This highly non-linear feedback potentially means that
noise may significantly affect the chemotactic response.

Previous studies have shown that, in the absence of chemoattractant gradients,
noise in (de)methylation reactions can lead to a power-law distribution of run
durations \cite{Korobkova04,Tu05}. The resulting super-diffusive motion may
enhance search efficiency compared to Brownian motion
\cite{Korobkova04,Matthaus09}. It has also recently been shown that slow
fluctuations in the methylation dynamics can allow for enhancement of drift in
linear gradients \cite{Matthaus09,Emonet08} at the expense of the ability of
cells to localise in regions of high ligand concentration \cite{Matthaus09}.
However, the mechanism by which noise enhances drift remains unclear.

In this paper we study the effects of receptor (de)methylation noise on
chemotactic performance. We show that below a threshold noise level the
steady-state performance of cells in a sinusoidal ligand field does not improve
as noise is reduced. We also find that an optimal noise level, comparable to
this noise threshold, maximises the drift velocity in exponential gradients.
Analytic approximations of a minimal model reveal that in the relevant regime
where motor switching is fast compared to adaptation, drift enhancement results
from the interplay of noise with the response of the motor switching rate to the
output of the signalling network.

To study the effects of signalling noise on the chemotactic behaviour of 
{\em E.~coli} we performed simulations of bacterial populations using a scheme
coupling swimming and signalling \cite{Jiang10}. Signalling dynamics are
simulated according to a recently-developed model \cite{Tu08,Shimizu10}
describing the CheA activity, receptor methylation level, and phosphorylated
CheYp level. The stochastic switching of the two-state motor with
CheYp-dependent switching rates, and the run-and-tumble dynamics of the
bacterium in a three-dimensional environment are also simulated. Noise was
introduced into the deterministic model considered in \cite{Jiang10} by adding a
Gaussian white noise source to the dynamics of methylation. The strength of this
noise is proportional to a parameter $\gamma^{-1}$ which is used to control the
impact of noise without changing the response time of the network (full details 
of the model can be found in the Supplementary Information).

\begin{figure}[tb]
	\includegraphics{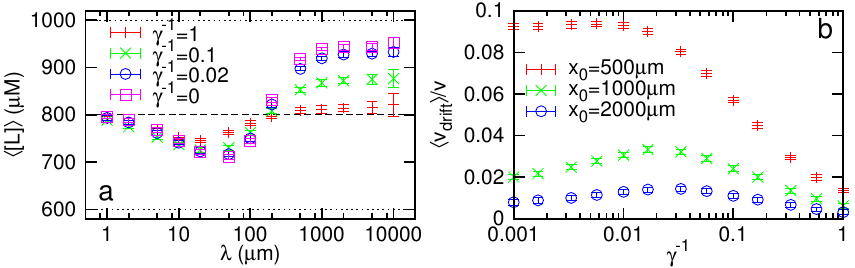}
	\caption{ \label{fig:full}
		(a) Mean steady-state ligand level $\avg{[L]}$ for a population of cells in 
			a sinusoidal ligand profile with $[L_0]=800\mu M$ and $A=0.25$, for 
			different noise strengths $\gamma^{-1}$. Dotted lines show the maximal and
			minimal concentrations at $[L]=[L_0](1\pm A)$. For short wavelengths 
			$\lambda\lesssim150\mm$ chemotactic cells perform worse than a simple 
			random walk, for which $\avg{[L]}=[L_0]$ (dashed line). Increasing the
			noise reduces the ability of cells to localise in regions of high ligand 
			concentration.
		(b) Transient drift velocity in an exponential ligand profile, with
			$[L_0]=20\mu M$ as the noise strength $\gamma^{-1}$ is varied. In shallow
			gradients $x_0>500\mm$ drift is fastest at intermediate noise strengths.
	}
\end{figure}

We first considered the ability of cells to localise in the vicinity of ligand
concentration peaks by studying the steady-state distribution of cells in a
sinusoidally-varying ligand profile, $[L(\mathbf{x})]= [L_0](1+A\cos2\pi
x/\lambda)$. Figure~\ref{fig:full}a shows the average ligand level
$\avg{[L]}=N_{\rm cells}^{-1}\sum_{i\in {\rm cells}} [L({\mathbf x}_i)]$
experienced by the simulated population as the gradient length-scale $\lambda$
and noise strength $\gamma^{-1}$ are varied. Non-chemotactic cells performing an
unbiased random walk would have $\avg{[L]}=[L_0]$. We first consider cells with
deterministic signalling, $\gamma^{-1}\to0$; the trajectories of such cells are
still noisy since motor switching and reorientation angles during tumbles remain
stochastic. For gradient length scales shorter than the typical run length of
unstimulated cells, $\lambda\lesssim5\mm$, cells are unable to react to the
extremely rapidly changing ligand level and effectively perform an unbiased
random walk. For long length scales $\lambda\gtrsim200\mm$ cells are able to
reliably localise in the vicinity of ligand concentration maxima. However, in an
intermediate range of length scales we find that $\avg{[L]}<[L_0]$. Here,
methylation is too slow to keep the CheYp level in the sensitive range of the
motor as cells run in directions of increasing ligand, and as a result cells
repeatedly overshoot ligand concentration peaks, an extreme example of the
`volcano effect' which has been observed previously \cite{Bray07}.

The shape of $\avg{[L]}$ as a function of $\lambda$ is largely unchanged as the
methylation noise strength $\gamma^{-1}$ is increased; however, the difference
between $\avg{[L]}$ and $[L_0]$ at a given $\lambda$ is decreased. Noise moves
cells away from concentration maxima when $\lambda$ is large, but also out of
minima when $\lambda\lesssim200\mm$.  However, since chemotaxis remains
detrimental in rapidly-varying profiles, it appears unlikely that signalling is
optimised for this type of environment. We therefore focus on the regime of long
wavelengths.  Importantly, here we find that there is an effective noise
threshold around $\gamma^{-1}\sim0.01$; provided signalling noise is kept below
this level, it is not detrimental for localisation near ligand maxima, which is
instead limited by the intrinsically random run-and-tumble motion of the cell.

Noise reduction in biochemical signalling is typically energetically costly,
requiring for example increased protein production or more rapid turn-over. Our
observation of a noise threshold suggests that there is no benefit, at least in
terms of localisation performance, to reducing noise below this level. Moreover,
it raises the possibility that this noise could somehow be exploited by the
cell. Bacterial chemotaxis has two conflicting goals \cite{Clark05}:
localisation in regions of high chemoattractant, and rapid drift in favourable
directions. The observed chemotactic response has been interpreted as a
compromise between these two objectives \cite{Clark05,Celani10}. It is therefore
important to also consider the effects of noise on the transient drift rate of
cells.

We therefore investigated transient chemotactic drift in an exponential ligand
gradient, $[L(\mathbf{x})]=[L_0]\exp(x/x_0)$.  Figure~\ref{fig:full}b shows the
drift velocity, estimated from the linear regime of the mean $x$-position
$\avg{x(t)}=N_{\rm cells}^{-1} \sum_{i\in{\rm cells}}x_i(t)$ as a function of
elapsed time, for a population initially located at $\mathbf{x}=\vec{0}$. We
considered only shallow gradients, $x_0\geq500\mm$, for which $\avg{x(t)}$
reaches a stable constant drift regime before saturation of ligand binding.
Based on experimentally-observed ramp responses \cite{Shimizu10}, cells are
expected to be sensitive to gradients with $x_0$ beyond $10000\mu m$.
Interestingly, we see that the drift velocity has a non-monotonic dependence on
methylation noise strength: drift in a cell with noisy methylation dynamics can
be faster than in cells with no methylation noise, and there is a
steepness-dependent noise strength that maximises the drift velocity. This
effect is also observed taking other measures of drift performance such as the
maximal drift velocity or the mean position of the population after some fixed
time. Our results are consistent with previous reports that cells with
signalling noise drift more rapidly in linear gradients
\cite{Matthaus09,Emonet08}. We note that the optimal noise strength
$\gamma^{-1}\approx0.01-0.02$ gives rise to a coefficient of variation in the
CheYp level of around $0.2$, comparable to the variability required \cite{Tu05}
to reproduce the experimentally-observed power law run-time distribution
\cite{Korobkova04}. Importantly, we also see that at the optimal noise level
neither the drift velocity in steep gradients nor the steady-state localisation
performance of cells is significantly compromised.

To understand the origin of this noise-induced drift enhancement in shallow
gradients we studied a minimal model of chemotactic drift, shown in
Fig.~\ref{fig:model}a (see the Supplementary Information for derivation). We
consider motion in one spatial dimension; cells can be in two states, `$+$' or
`$-$', which correspond to motion with velocity $\tilde{v}=\pm1$ in directions
of increasing or decreasing ligand concentration respectively.  We assume that
directional changes are instantaneous and hence no tumbling state is considered.
Furthermore we assume that the CheYp level simply tracks CheA activity, such
that the internal dynamics of the signalling pathway can be represented by a
single state variable, $\delta a(t)$, that represents the deviation of the
pathway activity from its adapted value (but with the direction of action of the
gradient reversed) and evolves according to $\dot{\delta a}=\tilde{v}r-\delta a+
\sigma\Gamma(t)$. Here $r$ represents the stimulus from the ligand gradient,
$\sigma$ is the strength of methylation noise, and $\Gamma(t)$ is a Gaussian
white noise process with unit variance. We take $r$ to be a constant since in
the full chemotaxis network the effective stimulus strength goes as
$\inlinepdiff{\log[L(x)]}{x}\sim x_0^{-1}$ for an exponential gradient.  Since
the noise-enhancement of drift is observed in shallow gradients we focus on the
regime of small $r<1$, a range of stimuli comparable in the rescaled parameter
space to the gradients for which drift enhancement is observed in the full
model. Finally, the switching propensity between `$+$' and `$-$' states is given
by $\omega(\delta a)=\kappa(1-\delta a)$ for $\delta a\leq1$ and $\omega(\delta
a)=0$ otherwise. This form for $\omega$ approximates the highly non-linear
dependence of the motor switching propensity on the CheYp level \cite{Cluzel00}
with $\kappa$ setting the typical rate of reorientations relative to adaptation,
and is chosen for computational simplicity. However, similar results are
observed with $\omega$ a smooth decreasing function of $\delta a$.  Following
\cite{Erban04}, evolution equations can be written for the joint probability
$p_\pm(\delta a;t)$ of a cell to be in the `$+$' or `$-$' state with internal
variable $\delta a$,
\begin{equation} \label{eq:pa}
	\inlinepdiff{p_\pm}{t}-
		\inlinepdiff{\left[\left(\delta a\mp r\right)p_\pm\right]}{\delta a}
	=\frac{\sigma^2}{2}\inlinepdiffn{p_\pm}{\delta a}{2}\mp
		\omega(\delta a)(p_+-p_-)
\end{equation} 
where the first term on the right-hand side of Eq.~\ref{eq:pa} is due to the
noise in the dynamics of $\delta a$. Thus in each state, cells effectively
diffuse in a potential $V_\pm(\delta a)\sim(\delta a\mp r)^2$ with diffusion
constant $\sigma^2/2$. The net population drift velocity in this minimal model
is given simply by $\avg{J(t)}=\int_{-\infty}^{\infty}\left[p_+(\delta
	a;t)-p_-(\delta a;t) \right]d\delta a$.

\begin{figure}[tb]
	\includegraphics{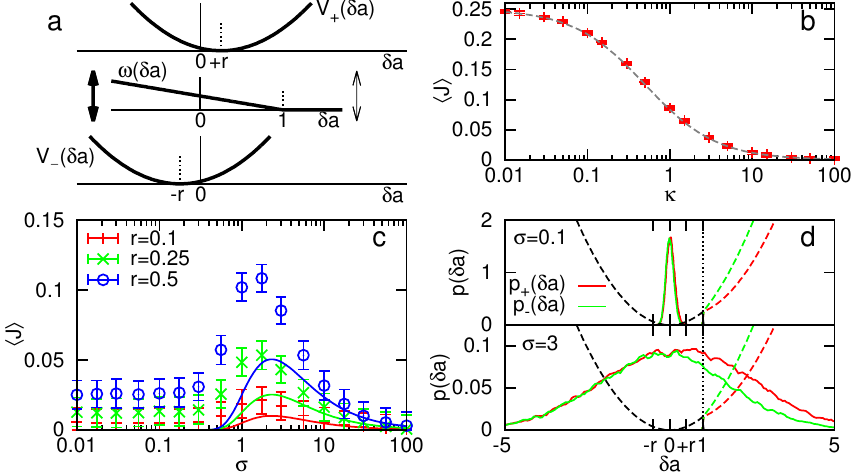}
	\caption{ \label{fig:minimal} \label{fig:model}
		(a) Schematic depiction of the minimal model. Cells switch between the two
			states `$+$' and `$-$' with rate $\omega(\delta a)$. The internal
			coordinate $\delta a$ diffuses within the potential $V_\pm(\delta a)$.
		(b) Steady-state drift $\avg{J}$ as switching rate $\kappa$ is varied, for
		  $r=0.25$ and $\sigma=0$. Points show simulation results; dashed line shows
		  $\avg{J}=r/(1+2\kappa)$.
		(c) Steady-state drift $\avg{J}$ as the noise strength $\sigma$ is varied
		  for different stimulus strengths $r$, with $\kappa=10$.  For small $r$,
		  $\avg{J}$ is maximised for $\sigma>0$. Lines show the approximate results
		  for $\kappa\to\infty$, Eq.~\ref{eq:large_k}.
		(d) Probability densities $p_\pm(\delta a)$ (solid lines) when $r=0.5$, with
		  $\kappa=10$. For small $\sigma$, $p_\pm(\delta a)$ are similar and peaked
		  around $\delta a=0$. For larger $\sigma$, the different forces experienced
		  when $\omega(\delta a)$ is small lead to differences between the
		  distributions $p_\pm(\delta a)$. Dashed lines show the effective 
			potentials $V_{\rm eff}(\delta a<1)$ and $V_\pm(\delta a>1)$.
		}
\end{figure}

In the absence of noise, $\sigma=0$, Eq.~\ref{eq:pa} can be solved to find the
steady-state drift. While the full expression is uninformative (see
Supplementary Information), for small $r\ll1$ it can be approximated as
$\avg{J}\approx r/(1+2\kappa)$ (see Fig.~\ref{fig:minimal}b). We see that
increasing $\kappa$ decreases $\avg{J}$, a result that also holds when
$\sigma>0$; intuitively, if the typical run duration is shorter, less
information can be extracted about the current direction during a single run,
and hence the reliability with which the tumbling propensity can be regulated is
reduced.

An exact solution to Eq.~\ref{eq:pa} when $\sigma>0$ is less straightforward.
Since in the full chemotaxis network the typical switching rate is fast compared
to the adaptation time, with an effective value of $\kappa\approx10$, we focus
in the remainder of the paper on the regime $\kappa\gg1$ for which both
numerical solutions and an analytic limit solution are possible.
Figure~\ref{fig:minimal}c shows the numerically-evaluated $\avg{J}$ at
$\kappa=10$ as the noise $\sigma$ is varied for different stimulus strengths
$r$. We can see that the minimal model qualitatively reproduces the results of
the full model for shallow gradients, with $\avg{J}$ showing a maximum at
$\sigma\gtrsim1$.

To understand the origin of this noise-induced maximum it is useful to consider
the limit of rapid switching, $\kappa\to\infty$. In the region $\delta a<1$,
cells rapidly exchange between `$+$' and `$-$' states, such that $p_+(\delta
a)=p_-(\delta a)$. In this region, therefore, cells spend equal time in each of
the two states while moving in the effective mixed potential $V_{\rm eff}(\delta
a)=[V_+(\delta a)+ V_-(\delta a)]/2\sim(\delta a^2+r^2)$. However, any cell
which crosses into the region $\delta a>1$, where $\omega(\delta a)=0$, will
experience only the potential associated with its current state, $V_\pm(\delta
a)$, until returning to the boundary $\delta a=1$. The average net drift can be
calculated in terms of the mean time spent in each region, $\delta a<1$ and
$\delta a>1$ in either the `$+$' and `$-$' states as
\begin{equation} \label{eq:J_times}
	\avg{J}=\frac{T^{\delta a>1}_+-T^{\delta a>1}_-}{T^{\delta
	a>1}_++2T^{\delta a<1}+T^{\delta a>1}_-}
\end{equation}
where $T^{\delta a<1}$ is the typical time spent in the region $\delta a<1$
accounting for both the `$+$' and `$-$' states and we have used the fact that
these cells have no bias in their direction. Evaluating the typical times spent
diffusing in the appropriate potentials we find 
\begin{equation}	\label{eq:large_k}
	\avg{J}\approx r\left[\frac{\exp({\sigma^{-2}})}{\sqrt{\pi\sigma^2}}
		-\frac{{\rm erfc}{(\sigma^{-1})}}{\sigma^2}\right],
\end{equation}
which can be seen to have a maximum at $\sigma\approx2$
(Fig.~\ref{fig:minimal}c, see Supplementary Information for an exact
expression and full derivation).

When the noise level is small, $\sigma<1$, cells occupy the minimum of $V_{\rm
eff}(\delta a)$ at $\delta a=0$ and $p_+(\delta a)\approx p_-(\delta a)$ (see
Fig.~\ref{fig:minimal}d), such that the net drift $\avg{J}\approx0$. As the
noise is increased, so too is the rate at which cells reach the transition point
$\delta a=1$, beyond which $\omega(\delta a)=0$. Importantly, the offset between
the minima of $V_+(\delta a)$ and $V_-(\delta a)$ means that cells in the region
$\delta a>1$ in the `$-$' state will experience a stronger force in the $-\delta
a$-direction than cells in the `$+$' state, $\abs{\inlinepdiff{V_+}{\delta a}}<
\abs{\inlinepdiff{V_-}{\delta a}}$. Hence the mean time spent diffusing in the
region $\delta a>1$ before returning to the boundary at $\delta a=1$ will be
longer in the `$+$' state than in the `$-$' state, $T^{\delta a>1}_+>T^{\delta
a>1}_-$. This is the origin of the drift enhancement by noise: as $\sigma$ is
increased beyond unity, cells spend an increasing amount of time in the region
$\delta a>1$, and so the magnitude of this effect and hence the drift $\avg{J}$
increase. For even larger values of the noise, the difference in the amount of
time spent in the `$+$' and `$-$' states decreases again, because now diffusion
dominates over the difference in forces. Hence $\avg{J}$ decreases again for
$\sigma\gg1$.

While Fig.~\ref{fig:minimal}c shows qualitative agreement between
Eq.~\ref{eq:large_k} and the numerical results for $\kappa=10$, $\avg{J}$
calculated for $\kappa\to\infty$ underestimates the drift when
$\kappa$ is finite (see also Supplementary Information). With a finite switching
rate, $p_\pm(\delta a<1)$ need not be precisely identical; indeed, an effective
positive feedback acts on differences between $p_\pm(\delta a<1)$ due to the
variation of $\omega(\delta a)$ with $\delta a<1$. As $\delta a$ becomes larger
and $\omega(\delta a)$ decreases, cells will spend more time in the potential
$V_\pm(\delta a)$. Since cells in the `$+$' state tend to drift towards larger
values of $\delta a$ than cells in the `$-$' state, cells will typically remain
in the `$+$' state for longer, allowing for further drift and amplifying the
differences between $p_+(\delta a)$ and $p_-(\delta a)$. This means that (i)
cells with $\delta a<1$ also contribute positively to $\avg{J}$; and (ii) more
cells enter the region $\delta a>1$ in the `$+$' than the `$-$' state, further
increasing $\avg{J}$.

\begin{figure}
	\includegraphics{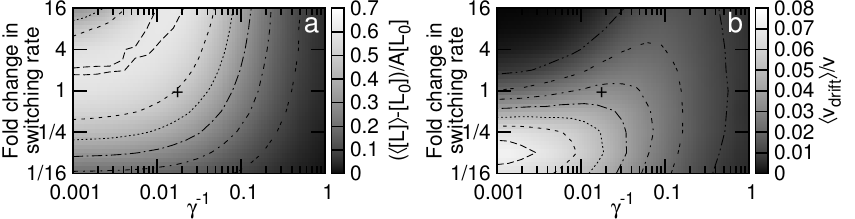}
	\caption{ \label{fig:vary_switch_rate}
		Effects of varying the noise strength $\gamma^{-1}$ and typical rate of
		motor switching in the full model on (a) steady-state ($\lambda=500\mu m$),
		and (b) transient ($x_0=1000\mu m$) chemotactic performance. At intermediate
		switching rates, the optimal drift velocity can be achieved with a finite
		noise level without compromising steady-state performance. The estimated
		wild-type parameters are indicted with $+$.
	}
\end{figure}

The maximal drift at the optimal noise strength for large $\kappa\gg1$ remains
less than $r$, the drift which is achieved for small $\kappa\to0$ and
$\sigma\to0$ (see Fig.~\ref{fig:minimal}b), suggesting the {\em E. coli} could
instead enhance chemotactic performance simply by reducing the switching rate
and signalling noise. However, chemotactic performance also depends on
steady-state localization. This motivated us to study the effects of varying the
rate of motor switching in the full model. As shown in
Fig.~\ref{fig:vary_switch_rate}, reducing the switching rate increases the
transient drift velocity at low noise levels, but this is accompanied by a
decrease in steady-state performance. This data emphasises that the choice of
chemotactic network parameters entails a trade-off between steady-state and
transient performance. The switching rate cannot be significantly decreased
without severely compromising steady-state performance. But at the switching
rates observed in real cells, a moderate level of noise improves drift
performance compared to a system without noise, without harming steady-state
localisation and additionally reducing the cost of signalling.

The principal result of our manuscript is an explanation of the mechanism by
which noise can enhance chemotactic drift. This mechanism is unlike the
noise-driven motion of Brownian ratchets or motors \cite{Reimann02} in that
drift reflects differences in the mean occupancy of internal states of the
system, rather than motion in an underlying spatial potential. The effect
described here can potentially enhance the bias of any two-state system in which
the rate of switching between these states is dependent on internal variables of
the system and is fast compared to the relaxation in either state. The essence
of the mechanism is that the fast switching obscures the difference between the
two states, unless noise is sufficiently strong that the system transiently
passes into a regime of slow switching. Then the asymmetry between the two
states is felt, and a bias is induced in the steady-state occupancy of the
states. This effect also differs from stochastic resonance \cite{Gammaitoni98}
since it is the enhancement by noise of a steady-state response to a constant
input, rather than the enhancement of a dynamic response to a time-varying
external signal.

Additionally, we have seen that a finite level of signalling noise can enhance
transient chemotactic drift in shallow gradients while not significantly
compromising drift in steep gradients or the ability to localise near
concentration peaks. More generally, our results suggest that cells may be able
to exploit intracellular signalling noise to improve behavioural responses.

\begin{acknowledgments}
We thank Andrew Mugler for comments on the manuscript. This work is part of the
research program of the ``Stichting voor Fundamenteel Onderzoek der Materie
(FOM)'', which is financially supported by the ``Nederlandse organisatie voor
Wetenschappelijk Onderzoek (NWO)''.
\end{acknowledgments}

\section{Supplementary Information for ``Signalling noise enhances
	chemotactic drift of \emph{E.~coli}''}
\setcounter{figure}{0}
\renewcommand{\thefigure}{S\arabic{figure}}
\setcounter{equation}{0}
\renewcommand{\theequation}{S\arabic{equation}}
\setcounter{table}{0}
\renewcommand{\thetable}{S\arabic{table}}

\section{Description of the full chemotaxis model} \label{sec:app}

Here we outline the full model of the chemotaxis pathway, adapted from
\cite{Jiang10}, used in the simulations. Parameter values used in the
simulations are listed in Table \ref{tab:params}.

It is assumed that the probability that a given receptor complex is in the
active state is in quasi equilibrium and is given by
\begin{equation} \label{eq:a}
	a=\frac{1}{1+\exp(N\epsilon(m,[L]))},
\end{equation}
where $\epsilon$ is the free-energy difference between the active and
inactive states and $N$ is the number of responding receptor dimers. The
free energy $\epsilon(m,[L])=f_m(m)+f_{[L]}([L])$ is a sum of contributions
dependent on the receptor methylation state $f_m(m)={\alpha}(m_0-m)$ and the
ligand concentration
$f_{[L]}([L])=-\log\left(\frac{1+[L]/K_A}{1+[L]/K_I}\right)$, where $K_A$
and $K_I$ are the dissociation constants of ligand to the active and
inactive receptors respectively, and roughly define the range of ligand
concentrations to which the cell can adapt.

The dynamics governing receptor methylation and demethylation is given by the
stochastic equation
\begin{equation} \label{eq:m}
	\diff{m}{t}=k_R(1-a)-k_Ba + \eta(t).
\end{equation}
Here we have introduced the Langevin noise term $\eta(t)$ to represent noise in
receptor (de)methylation reactions, with $\avg{\eta}=0$ and
$\avg{\eta(t)\eta(t')}=\gamma^{-1}(k_R(1-\overline{a}) + k_B\overline{a})
\delta(t-t')$. The values of the rate parameters $k_R$ and $k_B$ are set from
the experimental measurements \cite{Shimizu10} of the adapted kinase activity
$\overline{a}=k_R/(k_R+k_B)=0.5$ (we shall henceforth denote values of the
signalling variable in the adapted state with an overbar) and the slope of the
feedback transfer function
$\{\diff{}{a}(\diff{m}{t})\}_{a=\overline{a}}\approx-(k_R+k_B)=-0.03s^{-1}$.
Since the noise $\eta(t)$ is purely additive it does not alter the response time
of $m(t)$, which is set by $k_R$ and $k_B$. The parameter $\gamma^{-1}$ is
expected to scale with the number or receptor clusters as $N_{\rm rec}^{-1}$
\cite{Emonet08}, while the precise value of $\gamma^{-1}$ will also depend on
the cell volume and non-diffusive transport rates of CheR and CheB enzymes.

To propagate the noise from receptor modification, we include explicitly the
dynamics for the fraction of CheY which is phosphorylated, $y_p$,
\begin{equation} \label{eq:Yp}
	\diff{y_p}{t}=k_Ya(1-y_p)-k_Zy_p,
\end{equation}
Here the first term represents the phosphorylation of CheY by active
receptor-CheA complexes, and the second term accounts for dephosphorylation of
CheYp by the phosphatase CheZ. The fraction of CheY which is phophorylated in
adapted cells is
$\overline{y_p}=k_Y\overline{a}/(k_Y\overline{a}+k_Z)\approx0.3$
\cite{Alon98}.  Note that we do not introduce an additional source of noise
from the CheYp (de)phosphorylation reactions. This is because we expect such
fluctuations to be fast compared to methylation noise, and of small amplitude
due to the relatively high copy numbers of CheY and CheZ, and hence to have a
small impact on swimming behaviour.

\begin{table}
\begin{tabular}{ccc}
\toprule
Parameter & Value & Source\\
\colrule
$N$ & 6 & \cite{Jiang10} \\
$\alpha$ & 1.7 & \cite{Jiang10}\\
$m_0$ & 1 & \cite{Jiang10}\\
$K_A$ & $3000 \mu M$ & \cite{Jiang10}\\
$K_I$ & $18.2 \mu M$ & \cite{Jiang10}\\
$\overline{a}$ & 0.5 & \cite{Shimizu10} \\
$k_R$ & $0.015 \mbox{s}^{-1}$ & \cite{Shimizu10} \\
$k_B$ & $0.015 \mbox{s}^{-1}$ & \cite{Shimizu10} \\
$\overline{y_p}$ & 0.3 & \cite{Alon98} \\
$k_Z$ & $2 \mbox{s}^{-1}$ & \cite{Tu08}\\
$k_Y$ & $1.7 \mbox{s}^{-1}$ & calculated\\
$\overline{b}$& $0.25$ & \cite{Segall86} \\
${\tau}_0$ & 0.2 \mbox{s} & \cite{Jiang10,Alon98}\\
$H$ & 10 & \cite{Jiang10,Cluzel00}\\
$\beta$ & $282250 s^{-1}$ & calculated \\
$D_\theta$ & $0.123 {\rm rad}^2s^{-1}$ & \cite{Jiang10,Berg_book} \\
$v$ &	$16.5\mm s^{-1}$ & \cite{Jiang10,Alon98} \\
\botrule
\end{tabular}
\caption[?]{Parameter values}
\label{tab:params}
\end{table}

The tumbling bias is a sigmoidal function of $y_p$ with a Hill coefficient
$H\approx10$ \cite{Cluzel00}. CheYp modulates only the probability of the
motor to switch from counter-clockwise to clockwise rotation, and does not
affect the rate of the reverse transition \cite{Alon98}. Hence the
propensities for a cell to switch from running to tumbling and vice versa
are given by \cite{Jiang10}, 
\begin{equation} \label{eq:switch_propensities}
	\omega_{r\rightarrow t}(y_p)=\beta y_p^H,\ \ 
	\omega_{t\rightarrow r}=\tau^{-1},
\end{equation}
where $\tau$ is the average duration of a tumbling event, which is
independent of $y_p$. The constant $\beta$ is set such that the clockwise
bias in the adapted state is $\overline{b}=\omega_{r\rightarrow
t}(\overline{y_p})/ (\omega_{r\rightarrow
t}(\overline{y_p})+\omega_{t\rightarrow r})=0.25$ \cite{Segall86}. Upon
switching from tumbling to running, a new orientation for the cell is chosen
randomly and independently of the previous run direction. Introducing
correlations between the directions of consecutive runs, as has been
observed experimentally \cite{Berg72}, does not significantly affect our
results.

Simulations are initialised with a population of 10000 cells located at
$\mathbf{x}=\vec{0}$ adapted to the local environment. Each cell initially
has a random orientation and motor state chosen according to the adapted
tumbling bias, $\overline{b}$. In a simulation step of length $\delta t$,
running cells move a distance $\delta t \cdot v$ in the direction of their
current orientation, where the swimming speed $v$ is taken to be constant;
tumbling cells do not move. During runs, the run direction is also perturbed
by rotational diffusion with diffusion constant $D_\theta$; a random angular
displacement is added to the orientation at each time step to account for
this. The internal state is also updated according to equations
(\ref{eq:a}-\ref{eq:Yp}), and the motor state switches with probability
$\delta t\cdot\omega_{r\rightarrow t}$ or $\delta t\cdot\omega_{t\rightarrow
r}$, given by Eq.~\ref{eq:switch_propensities}, as appropriate.

\section{Derivation of the minimal model}

Here we present the complete derivation of the minimal model and its parameters
in terms of those of the full model.

We start by considering the internal dynamics of the activity $a$. Taking the
time-derivative of Eq.~\ref{eq:a} and substituting in Eq.~\ref{eq:m} gives
\begin{widetext} \begin{equation}
	\diff{a}{t}=\pdiff{a}{m}\diff{m}{t}+\pdiff{a}{[L]}\diff{[L]}{t}
		= Na(1-a)\left[\alpha(k_R(1-a)-k_Ba+\eta)+
			\frac{K_I-K_A}{([L]+K_A)([L]+K_I)}\pdiff{[L]}{t}\right].
\end{equation} \end{widetext}
Making the approximation $K_I\ll [L]\ll K_A$, and using the identity
$\overline{a}=k_R/(k_R+k_B)$, this expression simplifies to
\begin{equation}
	\diff{a}{t}=Na(1-a)
		\left[\frac{\alpha k_R}{\overline{a}}(\overline{a}-a)+
			\alpha\eta-\frac{1}{[L]}\diff{[L]}{t}\right].
\end{equation}
To model the reduced swimming dynamics in one spatial dimension we write
$\diff{[L]}{t}=\pm v\diff{[L]}{x}\vert_{\rm eff}$, where $\vert_{\rm eff}$
denotes an effective steepness calculated by integrating $\abs{\diff{[L]}{x}}$
over a uniform distribution of run angles in three-dimensional space. For an
exponential gradient $[L(\mathbf{x})]=[L_0]\exp(x/x_0)$ we therefore have
\begin{equation} \label{eq:a_dynamics}
	\diff{a}{t}=Na(1-a)\left[\frac{\alpha k_R}{\overline{a}}
		(\overline{a}-a)+\alpha\eta\mp\frac{v}{2x_0}\right].
\end{equation}

Turning to the switching dynamics, we first neglect the presence of the tumbling
state since tumbling events are relatively short compared to runs. Then the rate
of changing direction is simply given by $\omega(y_p)=\omega_{r\to t}(y_p)/2$,
where the factor of $1/2$ appears because only half of tumbling events will lead
to a change of direction.  Next we make a quasi-steady-state assumption that,
since the dynamics of $y_p(t)$ is fast compared to $a(t)$,
$y_p(t)=a(t)/\left[a(t)+k_Z/k_Y\right]$. Substituting into $\omega(y_p)$ and
linearizing about the adapted value $\overline{a}$ leads to
\begin{align}\label{eq:switch_expansion}
	\omega(a)\approx\frac{\beta\overline{y_p}^H}{2}\left[1-
		(\overline{a}-a)\frac{Hk_Z}{\overline{a}(k_Y\overline{a}+k_Z)}\right].
\end{align}
We define the displacement variable $\delta a$ according to
Eq.~\ref{eq:switch_expansion} such that $\omega(\delta a)
=\beta\overline{y_p}^H(1-\delta a)/2$. Next we substitute the corresponding
expression for $a(t)$ into Eq.~\ref{eq:a_dynamics} and rescale the units of time
according to $\tilde{t}=tN\alpha k_R(1-\overline{a})$. Finally, assuming that
the noise and stimulus terms as small deviations of the same order as $\delta a$
and retaining only first-order terms in $\delta a$ yields
\begin{equation}\label{eq:minimal_lang}
	\diff{\delta a}{\tilde{t}}=\pm\frac{vHk_Z}
			{2\alpha k_R(k_Y\overline{a}+k_Z)x_0}
		-\delta a
		+\frac{Hk_Z}{k_R(k_Y\overline{a}+k_Z)}\eta.
\end{equation}
We identify the first term with $\tilde{v}r$ and the rescaled velocity
$\tilde{v}=\pm1$, such that for a given value of $r$ the equivalent value of
$x_0$ using the parameters of Table~\ref{tab:params} is $x_0\approx2200\mm/r$.
Similarly, we can use the expression for the methylation noise strength in the
full model to identify
\begin{equation}
	\sigma=\frac{Hk_Z}{k_Y\overline{a}+k_Z}
		\sqrt{\frac{2(1-\overline{a})}{\gamma k_R}}
\approx57\gamma^{-1/2}.
\end{equation}
With these parameters, the stimuli in the $r<1$ regime of the minimal model are
comparable to gradients in the range for which non-monotonicity in the drift
velocity is observed in the full model.  The optimal noise level in the minimal
models is somewhat smaller than for the full model ($\sigma\approx2$ is
equivalent to $\gamma^{-1}\approx0.0012$), suggesting that the linear form of
$\omega(\delta a)$, which decreases more rapidly than $\omega_{r\to t}$, leads
to a slight overestimate of the effect of noise.

It remains only to convert the mean switching rate into the rescaled time units,
$\kappa=\frac{\beta\overline{y_p}^H}{2N\alpha
	k_R(1-\overline{a})}\approx10$.
	
\section{Exact solution for $\sigma=0$}

The full solution of the minimal model in the absence of noise is
\begin{equation}  \label{eq:noiseless} 
	\avg{J}=\begin{cases} 1-(1-r)
	\frac{{}_1{\rm F}_1\left(\kappa(1+r),1+2\kappa;-4r\kappa\right)}
		{{}_1{\rm F}_1\left(\kappa(1+r),2\kappa;-4r\kappa\right)} & r<1 \\ 
		1 & r\geq1
	\end{cases},
\end{equation}
where ${}_1{\rm F}_1(c_1,c_2;x)$ is a Kummer hypergeometric function.

\section{Exact solution for $\kappa\to\infty$}

The exact solution to the minimal model in the limit $\kappa\to\infty$ can be
expressed in terms of the typical time spent in the domains $\delta a<1$, and 
$\delta a>1$ in the `$+$' and `$-$' states, according to 
\begin{equation}
	\avg{J}=\frac{T^{\delta a>1}_{+}-T^{\delta a>1}_-}
		{T^{\delta a>1}_{+}+2T^{\delta a<1}+T^{\delta a>1}_-}.
\end{equation}
To determine the typical time spent in the region $\delta a<1$, diffusing in the
potential $V_{\rm eff}(\delta a)=(\delta a^2+r^2)/2$, before reaching the
boundary at $\delta a=1$ we calculate the mean first-passage time from a
position $\delta a=1-\varepsilon$, where $\varepsilon$ is a small displacement,
to $\delta a=1$, using the standard result \cite{Gardiner_book} 
\begin{align} \label{eq:supp_T_left}
	T^{\delta a<1}& =\frac{2}{\sigma^2}
		\int_{1-\varepsilon}^{1}dx\ e^{2V_{\rm eff}(x)/\sigma^2} 
		\int_{-\infty}^{x}dy\ e^{-2V_{\rm eff}(y)/\sigma^2} \\
		& \xrightarrow{\varepsilon\to0}\varepsilon\sqrt{\frac{\pi}{\sigma^2}}
			e^{1/\sigma^2}\left(1+{\rm erf}\frac{1}{\sigma}\right),
		\label{eq:supp_T_left_limit}
\end{align}
where we have used the fact that the diffusion constant in the $\delta
a$-coordinate is $\sigma^2/2$. Similarly, the time spent in the $\delta a>1$
region starting from a position $\delta a=1+\varepsilon$ can be calculated using
the potentials $V_+(\delta a)=(\delta a-r)^2/2+r$,
\begin{align}
	T_+^{\delta a>1}& =\frac{2}{\sigma^2}
		\int_{1}^{1+\varepsilon}dx\ e^{2V_+(x)/\sigma^2}
		\int_{x}^{\infty}dy\ e^{-2V_+(y)/\sigma^2} \\
		& \xrightarrow{\varepsilon\to0}\varepsilon\sqrt{\frac{\pi}{\sigma^2}}
			e^{(1-r)^2/\sigma^2}{\rm erfc}\frac{1-r}{\sigma},
\end{align}
and $V_-(\delta a)=(\delta a+r)^2/2-r$,
\begin{align}
	T_-^{\delta a>1}& =\frac{2}{\sigma^2}
		\int_{1}^{1+\varepsilon}dx\ e^{2V_-(x)/\sigma^2}
		\int_{x}^{\infty}dy\ e^{-2V_-(y)/\sigma^2} \\
		& \xrightarrow{\varepsilon\to0}\varepsilon\sqrt{\frac{\pi}{\sigma^2}}
			e^{(1+r)^2/\sigma^2}{\rm erfc}\frac{1+r}{\sigma}.
\end{align}
The constant offsets to $V_\pm(\delta a)$ are included so that the potential
landscape is continuous at $\delta a=1$, but have no effect on the final result.
While each of the first-passage times vanishes for $\varepsilon\to0$, as the
population of excursions into the relevant domain becomes dominated by extremely
short trajectories which return to the boundary almost immediately, the ratios 
of these times remain well-defined. This is because the relative times spent in 
each domain are determined predominantly by long trajectories with macroscopic 
durations, rather than the increasing number of vanishingly-short trajectories.

\begin{figure}[t]
	\begin{center}
		\includegraphics{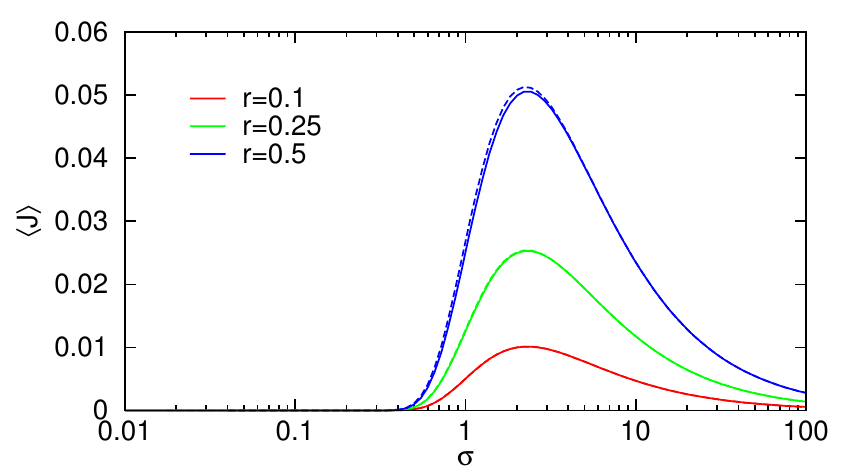}
	\end{center}
	\caption{The simplified Eq.~3 in the main text (full line) and the exact 
		result Eq.~\ref{eq:supp_kfast_full} (dashed) for $\kappa\to\infty$. For
		small $r\lesssim0.25$ the results are indistinguishable.}
	\label{fig:supp_kfast}
\end{figure}

Combining the results above leads to an exact solution for the drift in the
limit $\kappa\to\infty$,
\begin{equation} \label{eq:supp_kfast_full}
	\avg{J}=\frac{e^{-\frac{2r}{\sigma^2}}{\rm erfc}\frac{1-r}{\sigma}
		-e^\frac{2r}{\sigma^2}{\rm erfc}\frac{1+r}{\sigma}}
		{e^{-\frac{2r}{\sigma^2}}{\rm erfc}\frac{1-r}{\sigma}
			+2e^{-\frac{r^2}{\sigma^2}}\left(1+{\rm erf}\frac{1}{\sigma}\right)
		+e^\frac{2r}{\sigma^2}{\rm erfc}\frac{1+r}{\sigma}}.
\end{equation}
Figure~\ref{fig:supp_kfast} shows the excellent agreement between the full
result and the simplified Eq.~3 in the main text.

Importantly, Eq.~\ref{eq:supp_kfast_full} can also be derived by calculating the
equilibrium occupancy of the potentials according to the Boltzmann distribution.
We define the probability weight in the $\delta a<1$ region as 
\begin{equation}
	W^{\delta a<1}=\int_{-\infty}^{1}\delta a\ e^{-\beta V_{\rm eff}(\delta a)}
\end{equation}
with $\beta=2/\sigma^2$, and similarly for $W_\pm^{\delta a>1}$. By comparison
to Eq.~\ref{eq:supp_T_left}, it can then readily be seen that $W^{\delta
a<1}\sim T^{\delta a<1}/\varepsilon$, and the net drift follows
straightforwardly.


\begin{thebibliography}{10}

\bibitem{Berg72}
H.C. Berg and D.A. Brown.
\newblock {\em Nature}, 239:500--504, 1972.

\bibitem{Berg_book}
H.C. Berg.
\newblock {\em {\em E. coli} in Motion}.
\newblock Springer-Verlag, New York, 2004.

\bibitem{Segall86}
J.E. Segall, S.M. Block, and H.C. Berg.
\newblock {\em Proc. Natl Acad. Sci. USA}, 83:8987--8991, 1986.

\bibitem{Sourjik02}
V.~Sourjik and H.C. Berg.
\newblock {\em Proc. Natl Acad. Sci. USA}, 99:123--127, 2002.

\bibitem{Korobkova04}
E.~Korobkova, T.~Emonet, J.M.G. Vilar, T.S. Shimizu, and P.~Cluzel.
\newblock {\em Nature}, 428:574--578, 2004.

\bibitem{Shimizu10}
T.S. Shimizu, Y.~Tu, and H.C. Berg.
\newblock {\em Mol. Syst. Biol.}, 6:382, 2010.

\bibitem{Tu05}
Y.~Tu and G.~Grinstein.
\newblock {\em Phys. Rev. Lett.}, 94:208101, 2005.

\bibitem{Matthaus09}
F.~Matth\"{a}us, M.~Jagodi\v{c}, and J.~Dobnikar.
\newblock {\em Biophys. J.}, 97:946--957, 2009.

\bibitem{Emonet08}
T.~Emonet and P.~Cluzel.
\newblock {\em Proc. Natl Acad. Sci. USA}, 105:3304--3309, 2008.
M.W.~Sneddon, W.~Pontius, and T.~Emonet.
\newblock {\em Proc. Natl Acad. Sci. USA}, 109:805--810, 2012.

\bibitem{Jiang10}
L.~Jiang, Q.~Ouyang, and Y.~Tu.
\newblock {\em PLoS Comput. Biol.}, 6:e1000735, 2010.

\bibitem{Tu08}
Y.~Tu, T.S. Shimizu, and H.C. Berg.
\newblock {\em Proc. Natl Acad. Sci. USA}, 105:14855--14860, 2008.

\bibitem{Bray07}
D.~Bray, M.D. Levin, and K.~Lipkow.
\newblock {\em Curr. Biol.}, 17:12--19, 2007.

\bibitem{Clark05}
D.A. Clark and L.C. Grant.
\newblock {\em Proc. Natl Acad. Sci. USA}, 102:9150--9155, 2005.

\bibitem{Celani10}
A.~Celani and M.~Vergassola.
\newblock {\em Proc. Natl Acad. Sci. USA}, 107:1391--1396, 2010.

\bibitem{Cluzel00}
P.~Cluzel, M.~Surette, and S.~Leibler.
\newblock {\em Science}, 287:1652--1655, 2000.

\bibitem{Erban04}
R.~Erban and H.G. Othmer.
\newblock {\em {SIAM} J. Appl. Math.}, 65:361--391, 2004.

\bibitem{Reimann02}
P.~Reimann.
\newblock {\em Phys. Rep.}, 361:57--265, 2002.

\bibitem{Gammaitoni98}
L.~Gammaitoni, P.~H\"{a}nggi, H.~Jung, and F.~Marchesoni.
\newblock {\em Rev. Mod. Phys.}, 70:223--287, 1998.

\bibitem{Alon98}
U.~Alon, L.~Camarena, M.G. Suretter, B.A. Arcas, and Y.~and Liu.
\newblock {\em EMBO J.}, 17:4238--4248, 1998.

\bibitem{Gardiner_book}
C.W. Gardiner.
\newblock {\em Handbook of Stochastic Methods for Physics, Chemistry and the
	Natural Sciences, 2nd edition}.
\newblock Springer-Verlag, Berlin, 1985.
\end{thebibliography}
\end{document}